\documentstyle[aps,multicol,epsfig]{revtex}

\newcommand{\3}{$^3$He}
\newcommand{\tc}{T$_c$}

\newcommand{\rs}{$\rho_s$}
\newcommand{\rsb}{$\rho_s^b$}
\newcommand{\ea}{\emph{et al.}}

\begin{document}

\wideabs{

\title{Universal Behaviour of the Superfluid Fraction and \tc~
of \3 in Aerogel}

\author{G. Lawes$^a$, 
S.C.J. Kingsley$^a$\cite{byline}, 
N. Mulders$^b$, and J.M. Parpia$^a$}
\address{$^a$Laboratory of Atomic and Solid State Physics,
Cornell University, Ithaca, NY, USA.}
\address{$^b$Department of Physics and Astronomy, University
of Delaware, Newark, DL, USA}
\date{\today}
\maketitle

\begin{abstract}
We have investigated the superfluid transition of \3 in different
samples of silica aerogel.  By comparing new measurements
on a 99.5\% sample with previous observations 
on the behaviour of \3 in 98\% porous aerogel  we have found 
evidence for universal behaviour of
\3 in aerogel.  We relate both the transition temperature and superfluid 
density to the correlation length of the aerogel.
\end{abstract}

\pacs{PACS numbers: 67-57-z, 67.57 Pq}

}

The properties of bulk \3 are  well understood.  The extreme purity 
of \3 at low temperatures makes it an ideal system to study
the agreement between  theoretical and experimental results 
on non-conventional Cooper pairing in the absence of disorder. 
Disorder plays a crucial role in suppressing the pairing interaction
in high \tc~superconductors, the other well established non s-wave paired
system.
 The superfluid transition of \3 confined
to a sample of very porous silica aerogel was first reported four
years ago\cite{trey1,don1}.
  The aerogel provides a structural disorder background\cite{trey2}
to the liquid. 
  \3 is compressible, and the density
 can be continuously tuned by $\approx$30\%
  while maintaining a fixed disorder.
The \3 zero temperature coherence length $\xi_0$,
defined as $\xi_0=\hbar v_f / k_B T_c$,  varies from 180\AA~ to over 700\AA~
as a function of density.
Because the Cooper pairs in \3 form in a p-wave state, quasiparticle
scattering from the aerogel strands is pairbreaking\cite{note}.
Thus the \3-in-aerogel system is well suited to the
 exploration of 
the effect of impurity scattering and disorder 
on the superfluid transition and phase diagram.

The superfluidity of $^3$He in silica aerogel has been studied using 
torsional oscillators\cite{trey1,trey2,koichi,hook},
 NMR\cite{don1,hook,don2,barker,hook2} and 
sound propagation\cite{golov,matsubara}
 techniques.
These measurements  show that both the superfluid transition 
temperature (\tc) and superfluid density (\rs)
 of the \3 are suppressed
by the  disorder, but that the transition remains  sharp\cite{trey1}.
  This suppression 
is   sensitive to both the density and the microstructure of the 
aerogel sample. 

The simplest model for the effect of impurity scattering 
on the \3 superfluid transition is the homogeneous scattering
model (HSM) which is based on the Abrikosov-Gorkov model 
for a superconductor with magnetic impurities that induce 
pair-breaking  via spinflip scattering\cite{ag}.
This mechanism is similar to that  of diffuse
scattering of Cooper paired \3 from a surface\cite{adg},
 and is unable to explain the observed
behaviour.  Specifically, the observed suppression of the
superfluid density is much greater than predicted by this model.
  More sophisticated models, such as the 
isotropic inhomogeneous scattering model (IISM)\cite{erkki1,erkki2,erkki3}
proposed by Thuneberg and co-workers
 are able to quantitatively predict
the superfluid transition temperature of \3 in aerogel (for
small suppressions)  and  have had  success
 at qualitatively explaining the observed superfluid densities.
  
In this Letter we present data from several different experiments
on \3 in aerogel, including new results on \3 confined to a 99.5\%
porosity sample.  
This sample is a factor of four more dilute than any previously 
investigated\cite{note2} and  is crucial for understanding the
evolution from bulk \3 to a regime where impurity scattering
dominates.
  In comparing these different samples we find evidence
that the relation between superfluid density and the superfluid transition
temperature of \3 in aerogel follows a universal behaviour, independent
of the aerogel sample.  This is significant because both of these 
quantities are individually sensitive to the microstructure of the 
aerogel, and vary greatly from sample to sample.  We also present
evidence that the suppression of \tc~ can be related to  the correlation
length ($\xi_a$) of the aerogel sample.  
 
The aerogels used in the experiments discussed in this Letter were
grown under basic conditions\cite{schaefer,vacher,hasmy1,hasmy2}.
  Under these conditions gelation is
the result of diffusion limited aggregation of small ($\approx$30\AA~ 
diameter) primary silica particles.  The aerogels are
characterized by a fractal dimension (D$_f$) related to the real
space correlations, and a long length scale cutoff to these
correlations ($\xi_a$) above which the sample appears homogeneous.
The fractal exponent depends only on the gelation process, while the
cutoff length is also dependent on the average density.  Simulations
based on the diffusion-limited cluster-cluster aggregation (DLCA)
algorithm predict that the fractal exponent should lie between 1.7 and 
1.9, which is in good agreement with small-angle X-ray scattering 
(SAXS) measurements (Table~\ref{tab:agel_struct}).  
We note that $\xi_a$ in the most dilute sample, D, could not be inferred 
from the data as the SAXS did not extend to sufficiently small q.

\begin{table}[ht]
\begin{tabular}{rccc}
	& Porosity &	D$_f$ & $\xi_a$\\
Sample A & 0.98 & 1.9 & 1300\AA \\
Sample B & 0.98 & 1.8 & $\approx$ 900\AA \\
Sample C & 0.98 & 1.8 & 840\AA \\
Sample D & 0.995 & 1.7 & $\ge$2000\AA 
\end{tabular}
\caption{Parameters that characterize  aerogel samples.}
\label{tab:agel_struct}
\end{table}

Samples A and C have the same density, but were made under 
different conditions and have slightly different fractal dimension and
significantly different $\xi_a$.  Both samples A and C have been 
studied with SAXS.  Sample B was made under conditions very similar
to sample C.  We do not have direct information on its microstructure
from SAXS, but assume here that samples B and C are essentially identical.
The correlation length for sample D is within the range obtained from 
 simulations based on the DLCA algorithm and is consistent with the
SAXS data of Mulders on a 99.6\% sample ($\xi_a\ge$2000\AA).
  For a more extensive
discussion the reader is referred to reference \cite{trey2}.

The 99.5\% porosity aerogel used for our experiment was grown
inside the (on average) 100 $\mu$m large pores of a coarse 
 silver sinter.  Previous torsional oscillator experiments
have been affected by the presence of spurious resonances resulting from 
composite modes of \3 and aerogel whose frequency
 crosses the resonant frequency of the
cell\cite{trey1,hook}.
  The strength of these resonances grow as the
porosity of the aerogel sample increases, 
 affecting the quality of data\cite{trey1,koichi,hook}.
  In our cell the aerogel is clamped to the silver sinter, and the effect
of spurious resonances is strongly reduced.

We operated our torsional oscillator in self-resonant mode near 483 Hz.
  The temperature in the cell was measured using
a lanthanum diluted cerous magnesium nitrate
 ac susceptibility thermometer, thermally connected to the sample through 
a shared reservoir of \3. 
 Data was collected while the temperature increased at a 
 rate of 20 $\mu$K per hour. The period shift of the oscillator as 
the superfluid \3 decoupled from the torsion head provided both
the transition temperature of the \3 in aerogel  and the
superfluid density.

\begin{figure}[ht]
\vspace{-4mm}
\centerline{\epsfxsize=8cm \epsffile{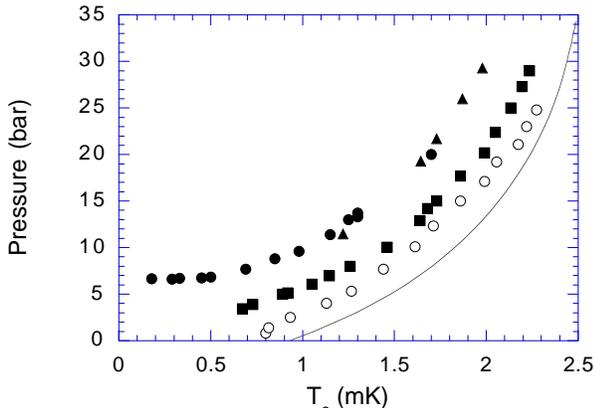}}
\caption{
The pressure is plotted versus \tc~for different aerogel samples.
The legend is as follows:  
 filled squares are cell A (98\% aerogel)\protect\cite{trey1}, 
 filled triangles are cell B (98\% aerogel)\protect\cite{trey2}, 
filled circles are for cell C (98\% aerogel)\protect\cite{koichi}
and empty circles are cell D (99.5\% aerogel).
The solid line is the superfluid transition curve
for bulk \3.}
\label{fig:tc}
\vspace{-5mm}
\end{figure}

Figure~\ref{fig:tc} shows the superfluid transition temperature for
several different aerogel samples.  Three of these samples (A, B and C)
have a nominal
porosity of 98\%, the fourth one (D) is our 99.5\% aerogel.
All of these measurements were done by monitoring the period
shift in torsional oscillators filled with the aerogel and \3.
  The difference
in transition temperature between samples A, B and C arises from 
differences in the microstructure of the aerogels (see 
table~\ref{tab:agel_struct} and reference~\cite{trey2}).  
The relative suppression of the transition temperature
(1-\tc/T$_{c0}$) (with T$_{c0}$ the transition temperature in bulk \3) is
larger at lower pressures (larger $\xi_0$) than at higer pressures
(smaller $\xi_0$).  In view of the fact that all the aerogels used in
these experiments have a similar fractal dimension and primary
particle size, one would expect them to be mutually self similar on
length scales shorter than the fractal cutoff length.
Thus for temperature dependent \3 coherence lengths $\xi(T)$ shorter
than $\xi_a$, the ratio of density of silica sampled at two different
$\xi(T)$ is independent of aerogel density.
  Experimentally one observes
a strong dependence of the suppression of \tc~on the aerogel density. 

\begin{figure}[ht]
\vspace{-4mm}
\centerline{\epsfxsize=8cm \epsffile{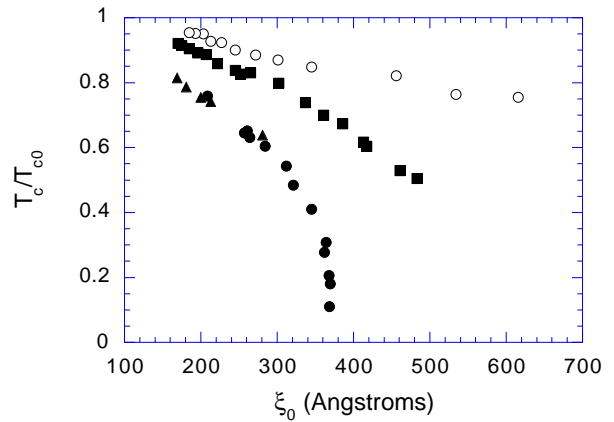}}
\caption{
The superfluid transition temperature for \3 in aerogel, scaled by
the bulk transition temperature, is plotted for several different
samples as a function of $\xi_0$.  The symbols are the same as shown in
Fig.~\ref{fig:tc}.}
\label{fig:xi}
\vspace{-5mm}
\end{figure}

In figure~\ref{fig:xi} we show the dependence of the relative suppression
of \tc~as a function of  $\xi_0$.  At short
coherence lengths, the relative suppression is small for all samples.
However, when the pressure is reduced, increasing $\xi_0$, this 
suppression shows a marked dependence on the microstructure
of the aerogel.
 The transition
temperature  shows an evolution from a strong impurity  scattering
regime when the \3 is confined to a 98\% porosity aerogel sample
 towards the behaviour of bulk \3 in the 99.5\% sample. 

\begin{figure}[ht]
\vspace{-4mm}
\centerline{\epsfxsize=8cm \epsffile{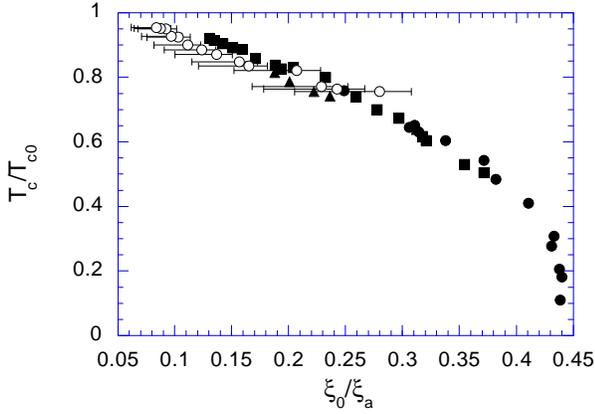}}
\caption{
Relative suppression of \tc~versus $\xi_0$/$\xi_a$.
The correlation length for the 99.5\% sample is taken to
be 2200\AA~with the error bars corresponding to the high
and low values of 3000\AA~and 2000\AA.}
\label{fig:xi_a}
\vspace{-5mm}
\end{figure}

Motivated by figure~\ref{fig:xi} we plot the relative suppression of 
\tc~against  $\xi_0$ scaled
by the aerogel correlation length $\xi_a$. 
The aerogel-limited mean free path ($l_g$) would be another
natural choice to compare against $\xi_0$.  However, this length
scale has not been independently measured.  If
we use the HSM to determine $l_g$ from \tc/T$_{c0}$, the values
of $l_g$ show a strong pressure dependence\cite{trey3}.
Figure~\ref{fig:xi_a} shows that the scaled transition temperature
depends solely on the ratio of $\xi_0$ to $\xi_a$ \emph{independent}
of the aerogel density.  The error bars for sample D result from the
high and low estimates (2000\AA~ and 3000\AA~ respectively) for the
correlation length for a 99.5\% porosity sample from DLCA simulations.

\begin{figure}[ht]
\vspace{-4mm}
\centerline{\epsfxsize=8cm \epsffile{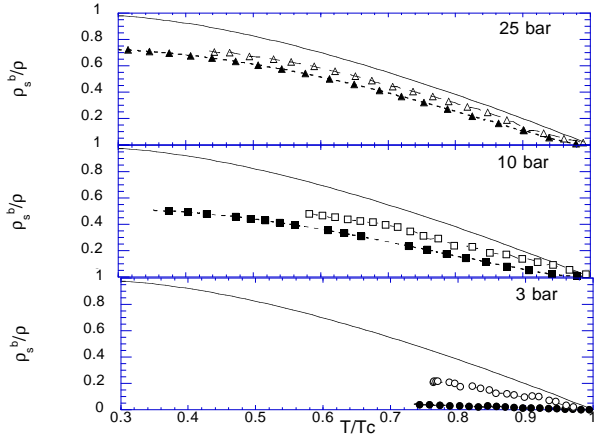}}
\caption{
We plot the bare superfluid density versus reduced temperature for 
cell A (98\%), filled symbols,  and cell D (99.5\%), empty symbols.}
\label{fig:rho}
\vspace{-5mm}
\end{figure}

We determined the temperature-dependent
 superfluid density of \3 in 99.5\% aerogel using
the shift in resonant frequency of our torsional oscillator upon
warming.  
A small, temperature dependent contribution due
to bulk \3 in the cell was  subtracted.
The remaining shift was scaled by the period
change due to filling the cell with  \3 at 50 mK and by the tortuosity 
 (measured with $^4$He) to obtain the superfluid density.
  Figure~\ref{fig:rho}
 shows the
bare superfluid density (\rsb/$\rho$) plotted versus the reduced
temperature T/\tc~for different aerogel samples and pressures.  The
bare superfluid density is obtained from  \rs~by stripping
away the Fermi liquid factor\cite{leggett} according to:

\begin{equation}
\frac{\rho_s^b}{\rho}
=\frac{(1+\frac{1}{3}F_1)\frac{\rho_s}{\rho}}
{1+\frac{1}{3}F_1\frac{\rho_s}{\rho}}
\end{equation} 

and is equivalent to 1-Y(T) where Y(T) is the temperature dependent
Yosida function for bulk \3.

As with the transition temperature \tc, the \rsb~for these
 two different samples are
similar at high pressures and are both close to the bulk value.  At the
lowest pressure, there is a factor of five difference in \rsb/$\rho$~
 at the
same reduced temperature between the 99.5\% sample and the 98\% sample;  
 \rsb/$\rho$~is more strongly suppressed by the aerogel than \tc.
There is also a large difference between the suppression factors of 
\tc~ and \rsb~ in the same aerogel sample.
  This large suppression in \rs~
with respect to \tc~is consistent with measurements made on other
 samples\cite{trey1,trey2}, and cannot be explained with homogenous scattering
models for \3 in aerogel\cite{erkki1}.

\begin{figure}[ht]
\vspace{-4mm}
\centerline{\epsfxsize=8cm \epsffile{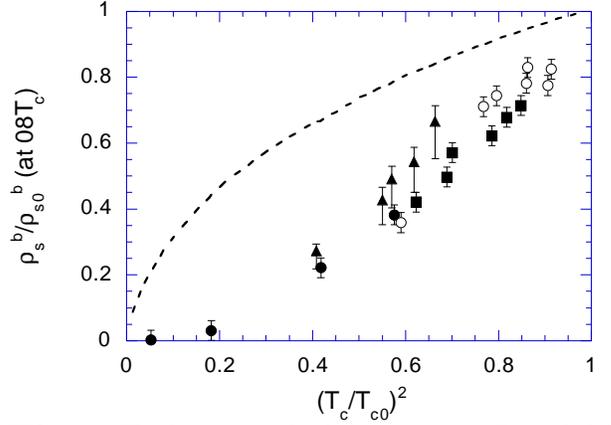}}
\caption{
The bare superfluid density at 0.8\tc~scaled by 
the bare bulk superfluid density
is plotted versus the square of the scaled \tc.  The symbols are
the same as in Fig.~\ref{fig:tc}.  The dashed line is the theoretical
curve from the HSM.}
\label{fig:rho2}
\vspace{-5mm}
\end{figure}

In order to better understand the superfluid transition of \3 in aerogel,
 in figure~\ref{fig:rho2}
 we plot \rsb/\rs~at 0.8\tc~against (\tc/T$_{c0}$)$^2$~
 for several  aerogel samples.  
The error bars shown for cell B arise from a spurious resonance that 
introduces uncertainty in the determination of $\rho_s$.
As in figure~\ref{fig:xi_a}, 
the data collapses on a  universal curve.
The dashed line is the prediction for a homogeneously scattering
model based on the Abrikosov-Gorkov equation\cite{erkki1}.
 This plot compares aerogels with \emph{different} 
densities---there is a factor of four difference in the average impurity
density between the 98\% samples and the 99.5\% aerogel sample.  
Furthermore, the coherence length of the Cooper pairs varies 
from 180 \AA~to
 600 \AA~over this data, yet the very strong pressure 
dependence shown in figures~\ref{fig:xi} and~\ref{fig:rho}
 has been factored out in this plot.  

  The
correlations in the aerogel will affect the suppression of \tc~and the 
evolution of \rsb~relative to a homogeous disorder, but this
suppression apparently  depends only on  the
correlation length of the sample and possibly the fractal exponent.
The steep slope of the data in figure~\ref{fig:rho2} is evidence that
the fractal nature of the aerogel plays an important role in
the development of $\rho_s$\cite{erkki4}. 
  Since all of these samples
were base-catalyzed, the behaviour of \3 in each aerogel
is determined mainly by $\xi_a$.
As long as $\xi_0$ is much less than $\xi_a$,
 the disorder sampled by the ensemble of Cooper pairs should be insensitive
to changes in the temperature-dependent coherence length
$\xi$(T), until $\xi$(T)$\approx\xi_a$.
That is, the system has a conformal symmetry normally
absent in disordered systems.  The evidence of this
one-parameter scaling is displayed in figure~\ref{fig:rho2};
 \rsb~is  a function
of T$_c$/T$_{c0}$ (or equivalently $\xi_0$/$\xi_a$) only.
  This behaviour is reminiscent of the
compilation of data from disordered high \tc~materials
  by Franz \emph{et al}\cite{franz} with the  exception
 that the relative suppression of \tc/T$_{c0}$ and
$\rho_s/\rho_{s0}$ in the high \tc~materials
do not follow a universal behaviour, presumably because
the impurities are not fractally correlated.

In order to compare the behaviour of \tc~and \rsb~of \3 in 
aerogel with \3 in bulk it will be necessary to understand precisely
\emph{how} the fractal disorder affects the superfluid pairing mechanism.
  One test for models of non-conventional Cooper pairing in the
presence of disorder would be
to predict the functional form of the universal curve for
\rsb~ versus (\tc)$^2$ for \3 in (base-catalyzed) aerogel. 
The IISM of Thuneberg \ea~ predicts a relation between 
$\rho_s^b/\rho_{s0}^b$ and T$_c$/T$_{c0}$ similar to
the trend illustrated in figure~\ref{fig:rho2},
 showing behaviour very different
than Abrikosov-Gorkov model\cite{erkki1,erkki2}.  This model does 
not explicitly consider the fractal nature of the aerogel, but 
shows how inhomogeneities in the disorder can lead to a large
suppression of \rsb~relative to \tc.

In this letter we have presented data from our measurements on \3
in a very dilute 99.5\% porous silica aerogel.  The values of \tc~and
\rsb~fall between those of \3 in bulk and \3 in denser aerogel samples.
We also present \tc~and \rsb~data  for \3 in aerogel
experiments performed at Cornell that show  universal behaviour
that can be traced to the fractal structure of the aerogel.  
In order to more fully understand this exciting physical system, 
more attention must be devoted to understanding the microstructure
of the aerogel. Specifically, the universal scaling discussed above
depends strongly on the fact that the fractal exponent real-space 
correlations is similar for all the aerogel samples.  It would be
interesting to study neutrally catalyzed silica aerogels, which
have a different fractal exponent than the base-catalyzed 
samples\cite{vacher}.  
 \3 would be expected to follow \emph{different} universal behaviour
when confined to base-catalyzed and neutrally catalyzed aerogels.  
As yet, there is no clear theoretical picture for why the transition
temperature \tc should depend on the ratio $\xi_0/\xi_a$. 
The explanation for this behaviour will provide insight into
how correlated disorder affects non-conventional Cooper pairing.

We would like to acknowledge helpful conversations with T.L. Ho, S. Yip,
E. Thuneberg, J. Beamish, A. Golov
 and M.H.W. Chan.  This research was supported
by the NSF under DMR-9705295.

%\begin{figure}
%\caption{The pressure is plotted versus \tc~ for different aerogel
%samples.  The filled circles, empty circles, filled diamonds and 
%crosses are for 98\% aerogel samples; the empty squares are for
%our 99.5\% porosity sample.  The solid line the the superfluid transition
%curve for bulk \3.}
%\label{fig:tc}
%\end{figure}

%\begin{figure}
%\caption{The superfluid transition temperature for \3 in aerogel, scaled 
%by the bulk transition temperature, is plotted for several different
%samples as a function of $\xi_0$.  The filled circles, filled diamonds and
%empty circles are 98\% samples; the empty squares are the 99.5\% aerogel
%sample.}
%\label{fig:xi}
%\end{figure}

%\begin{figure}
%\caption{We plot the superfluid density versus reduced temperature for two 
%different aerogel samples.  The filled symbols are for the 98\% sample, 
%the empty symbols are the 99.5\% sample.  The top graph is the superfluid
%density at 25 bar, the middle graph is at 10 bar and the bottom graph
%is at 2.5 bar for the 99.5\% and 3.4 bar for the 98\%. }
%\label{fig:rho}
%\end{figure}

%\begin{figure}
%\caption{The bare superfluid density, scaled by the bare bulk superfluid
%density is plotted versus the square of the reduced temperature for
%several different samples.  The filled circles, empty circles, filled
%diamonds and crosses are for 98\% samples, and the empty squares 
%are the 99.5\% sample.  The dashed line is a theoretical curve from
%the HSM which is based on Abrikosov-Gorkov theory.}
%\label{fig:rho2}
%\end{figure}


\begin{thebibliography}{99}
\bibitem{note}
Because the silica is much smaller in diameter than the coherence 
length, the impurities penetrate into the Cooper pairs, in contrast
to quasi-cylindrical pores with diameters $\approx\xi_0$ in which 
superfluidity is completely destroyed.
\bibitem{note2}
Experiments at Manchester \protect\cite{hook} on a 99\% sample have determined
the \tc~ of \3 in this dilute aerogel, but have not yet yielded information
on the superfluid density.
\bibitem{note4}
The $T=0$ value of $\rho_{s0}^b$ is 1, since the bulk Yosida function
is zero at $T=0$.
\end{thebibliography}
\end{document}